\documentclass{article}
\usepackage[margin=1in]{geometry}
\usepackage[T1]{fontenc}
\usepackage[utf8]{inputenc}
\usepackage{authblk} 
\usepackage{textcomp}
\usepackage{lmodern}
\usepackage{amsmath, amssymb}
\usepackage{graphicx}
\usepackage{enumitem}
\usepackage[numbers, sort&compress]{natbib}
\usepackage{hyperref}
\hypersetup{
    colorlinks=true,
    linkcolor=green,
    filecolor=magenta,      
    urlcolor=cyan,
    citecolor=blue,
}
\usepackage{tabularx, booktabs, array}
\usepackage{longtable}
\usepackage{multirow}
\usepackage{makecell}
\newcolumntype{L}{>{\raggedright\arraybackslash}X}
\usepackage{placeins}
\usepackage{float}
\usepackage{titlesec}
\titleformat{\section}{\normalfont\fontsize{12}{14}\bfseries\sffamily}{\thesection}{1em}{}
\titleformat{\subsection}{\normalfont\fontsize{11}{13}\bfseries\sffamily}{\thesubsection}{1em}{}
\titleformat{\subsubsection}{\normalfont\fontsize{10}{12}\bfseries\sffamily}{\thesubsubsection}{1em}{}
\usepackage{caption}
\providecommand{\keywords}[1]{%
  \par\noindent\textbf{Keywords:} #1%
}
\title{\bfseries A Survey on World Models Grounded in Acoustic Physical Information}

\author[1]{Xiaoliang Chen\thanks{Corresponding author: chenxiaoliang@soundai.com}}
\author[1]{Le Chang}
\author[1]{Xin Yu}
\author[1]{Yunhe Huang}
\author[1]{Xianling Tu}

\affil[1]{SoundAI Technology Co., Ltd.}

\date{}

\begin{document}

\maketitle

\begin{abstract}
This survey provides a comprehensive overview of the emerging field of world models grounded in the foundation of acoustic physical information. It examines the theoretical underpinnings, essential methodological frameworks, and recent technological advancements in leveraging acoustic signals for high-fidelity environmental perception, causal physical reasoning, and predictive simulation of dynamic events. The survey explains how acoustic signals, as direct carriers of mechanical wave energy from physical events, encode rich, latent information about material properties, internal geometric structures, and complex interaction dynamics. Specifically, this survey establishes the theoretical foundation by explaining how fundamental physical laws govern the encoding of physical information within acoustic signals. It then reviews the core methodological pillars, including Physics-Informed Neural Networks (PINNs), generative models, and self-supervised multimodal learning frameworks. Furthermore, the survey details the significant applications of acoustic world models in robotics, autonomous driving, healthcare, and finance. Finally, it systematically outlines the important technical and ethical challenges while proposing a concrete roadmap for future research directions toward robust, causal, uncertainty-aware, and responsible acoustic intelligence. These elements collectively point to a research pathway towards embodied active acoustic intelligence, empowering AI systems to construct an internal "intuitive physics" engine through sound.
\end{abstract}


\section{Introduction}

The long-standing quest for Artificial General Intelligence (AGI), which has seen tremendous progress with large-scale foundation models \cite{brown2020language, touvron2019open}, has increasingly converged on the need to construct intelligent agents capable of understanding, predicting, and meaningfully interacting with their physical environments. This paradigm marks a notable shift away from traditional approaches grounded solely in statistical pattern recognition and data-driven correlations—a powerfully articulated in the context of ungrounded language models \cite{bender2020climbing}—towards more sophisticated models that embody causal inference and internalize the physical laws governing our surroundings. Central to this transformation is the concept of \textit{world models}, internalized representations that are both generative and learnable. Such models enable agents to simulate their environments efficiently within compressed latent spaces, reason retrospectively and prospectively about events, and accurately predict outcomes resulting from their actions or from external changes and interventions \cite{ha2018world, hafner2019dream, lecun2022path, matsuo2022deep}.

Traditionally, visual perception, driven by breakthroughs in deep learning \cite{krizhevsky2017imagenet, dosovitskiy2020image}, has dominated the sensory modalities used to construct these internal models. However, despite its ubiquity and intuitive appeal, visual information alone imposes fundamental constraints. Visual sensing is inherently susceptible to practical limitations including poor illumination conditions, occlusions, atmospheric disturbances, and adverse weather scenarios, significantly hindering robustness in real-world applications \cite{hendrycks2019benchmarking}. More fundamentally, visual perception predominantly captures surface-level phenomena, frequently failing to convey intrinsic properties such as mass, internal composition, temperature gradients, structural integrity, and contact forces. These intrinsic characteristics are often ambiguous or entirely inaccessible through visual data alone, underscoring the need for complementary modalities capable of addressing these fundamental gaps.

In contrast, acoustic information provides a highly promising yet often overlooked complementary modality that can significantly augment and enrich world models. Sound, fundamentally, is a direct physical manifestation of mechanical interactions and events, arising from vibrations propagating through media and carrying explicit signatures of the underlying physical processes. Each acoustic signal encodes specific properties and dynamics of its originating phenomena—be it the resonant frequencies emitted by a struck object indicative of its material properties such as elasticity and density, or the characteristic turbulence sounds generated by fluids described by complex fluid dynamics. These acoustic signals inherently embed important information about material composition, internal structures, geometrical configurations, and dynamical interactions, deeply rooted in principles of auditory scene analysis \cite{bregman1994auditory, vesely1994computational}. Thus, by decoding acoustic signals through physical reasoning, intelligent systems can cultivate a nuanced "physical intuition," acquiring rich insights into the environment that are inherently complementary and orthogonal to visual perception.

This survey aims to provide a comprehensive and structured review of the emerging research field focused on acoustic-based physical world models. By integrating insights from acoustics, machine learning, signal processing, robotics, and cognitive science, we aim to present a comprehensive narrative highlighting theoretical underpinnings, methodological advancements, diverse application domains, and persistent challenges. We contend that interpreting sound through explicit physical frameworks significantly deepens AI's capacity for causal understanding, ultimately enhancing its perceptual precision, reasoning capabilities, and predictive accuracy. To achieve this, this survey first establishes the theoretical foundations of how acoustic physics encodes rich, causal, and interpretable information. Subsequently, it delves into the methodological pillars, emphasizing advanced techniques like Physics-Informed Neural Networks and generative modeling that bridge theoretical physics and data-driven AI. The survey then showcases the significant applications in key domains such as robotics and healthcare, demonstrating the practical impact of acoustically-grounded world models. Finally, it offers an important analysis of prevailing challenges and prospective research directions, examining technical and ethical issues while outlining a robust research pathway towards causal and adaptive acoustic intelligence.

Through this comprehensive review of existing state-of-the-art techniques, theoretical advancements, practical implementations, and forward-looking perspectives, we intend to foster deeper scholarly engagement and innovation in the field of acoustic world models. Ultimately, the insights presented aim to propel the development of next-generation AI systems endowed with a deep understanding of physical principles, capable of sophisticated environmental interaction, nuanced prediction, and genuinely intelligent, contextually aware decision-making.

\section{Theoretical Foundations: Acoustic Representation of Physical Laws}

Acoustic technology, characterized by its unique sensitivity to the physical environment, exceptional penetrability, and direct causal linkage to physical processes, constitutes an essential perceptual modality for a acoustic world model. Acoustic signals are not superficial background noise but are direct, wave-based encodings of physical events, material properties, and environmental structures. They are the radiated mechanical energy from dynamic phenomena, governed by the fundamental laws of mechanics and thermodynamics. The physical information they contain far surpasses what casual listening can perceive, providing AI with the deep physical insights necessary for "identifying objects by sound" and "understanding environments by sound," as envisioned by early pioneers in auditory scene analysis \cite{bregman1994auditory, vesely1994computational}. This section establishes the physical principles that enables this acoustic cognition.

\subsection{Estimating Physical Properties from Structure Vibrations}
The acoustic signature of a solid object is a direct manifestation of its vibrational behavior upon excitation. This behavior is governed by the laws of elastodynamics, and for a linear, isotropic, and homogeneous elastic solid in the absence of body forces, the displacement vector field $\mathbf{u}(x, t)$ is described by the generalized linear wave equation:
\begin{equation}
    (\lambda + \mu) \nabla(\nabla \cdot \mathbf{u}) + \mu \nabla^2 \mathbf{u} = \rho \frac{\partial^2 \mathbf{u}}{\partial t^2}
\end{equation}
where $\rho$ is the density, and $\lambda$ and $\mu$ are the Lamé parameters, which are functions of the material's Young's modulus $E$ and Poisson's ratio $\nu$. This equation supports two types of waves: longitudinal (pressure) waves and transverse (shear) waves, whose speeds are determined by these material properties. When an object is struck, it vibrates in a superposition of its natural modes or \textit{eigenmodes}. These are the special solutions to the wave equation subject to the object's specific geometry and boundary conditions (e.g., fixed, free, or simply supported). Each eigenmode, $\phi_n(x)$, has a corresponding natural frequency, or \textit{eigenvalue}, $\omega_n$. For example, the natural frequencies of a uniform beam undergoing transverse vibration are given by:
\begin{equation}
	\omega_n = (\frac{\beta_n}{L})^2 \sqrt{\frac{EI}{\rho A}}
\end{equation}
where $L$ is the length, $A$ is the cross-sectional area, $I$ is the area moment of inertia (a geometric property), and $\beta_n$ are constants determined by the boundary conditions.

\begin{itemize}
    \item Material and Modal Inversion: The sound radiated by a vibrating object is a summation of tones at these natural frequencies, $\omega_n$. The relative amplitudes and decay rates of these tones create the object's unique timbre. This provides a rich, high-dimensional acoustic fingerprint for inferring physical properties.
        \begin{itemize}
            \item Material Properties: As seen in the equations, the frequencies are directly tied to material parameters like $E$ and $\rho$. Therefore, by analyzing the spectrum of an impact sound, a model can work backward to infer these intrinsic material properties. This is the basis for a large body of work on material identification from sound \cite{traer2016statistics, giordano2003material}.
            \item Damping and Internal Friction: The rate at which each modal vibration decays is determined by the material's internal friction and energy radiation into the surrounding medium. This decay rate is quantified by the Quality factor (Q-factor) for each mode. A high Q-factor (e.g., in metals) leads to sustained, high-frequency, harmonic ringing, while a low Q-factor (e.g., in wood or plastic) leads to a rapidly decaying, dull thud. This temporal decay information is as important as the spectral content for material classification.
            \item Geometric Properties: The geometry of the object, encapsulated in terms like $L$, $A$, and $I$, along with its boundary conditions, dictates the specific set of eigenfrequencies. While inverting for the full 3D shape from a single sound is a highly ill-posed problem, machine learning models can learn to recognize the acoustic signatures of specific shapes or estimate key geometric parameters \cite{schwarz2000resonant}. Recent works have demonstrated that deep learning models can leverage these rich acoustic signatures to classify object materials with high accuracy \cite{jodhani2023ultrasonic}, estimate the viscosity of liquids (which affects the damping of a container's vibrations) \cite{park2022porosity}, or even infer the particle size of granular materials from the statistical properties of many small impacts during shaking \cite{guo2022deep}.
        \end{itemize}

    \item Internal Structure and Defect Detection: The penetrative power of sound waves makes them fundamental to Non-Destructive Testing (NDT) and Non-Destructive Evaluation (NDE) \cite{cook2002sound}. When propagating through an object, sound waves are reflected, refracted, and scattered at interfaces between different media, such as voids, cracks, or foreign inclusions. The acoustic impedance mismatch at these interfaces governs the reflection and transmission coefficients.
        \begin{itemize}
            \item Pulse-Echo and Time-of-Flight Analysis: In active sensing, an ultrasonic pulse is emitted, and the time-of-flight and amplitude of returning echoes are analyzed. This allows for the precise localization of internal boundaries and defects, forming the basis of ultrasonic imaging (known as B-scans).
            \item Acoustic Resonance Spectroscopy (ARS): This passive technique analyzes the complete resonant spectrum of an object under wide-frequency excitation. The presence of an internal flaw, even a microscopic one, alters the object's effective stiffness and mass distribution, causing a predictable shift in its resonant peaks. By comparing the measured spectrum to a reference spectrum of a "healthy" object, minute internal flaws can be detected \cite{avanzini2001controlling}. The application of deep learning to automate the analysis of these complex spectral shifts has greatly enhanced the efficiency of NDT in industrial settings, such as for identifying welding defects or composite material delamination \cite{bregnian1993auditory, wang2024multimodal}. This "acoustic X-ray" capability allows a world model to perceive the physical state of an object "beneath the surface," moving beyond mere appearance to internal integrity.
        \end{itemize}
\end{itemize}

\subsection{Decoding Acoustic Events from Contact to Aeroacoustics}
Nearly all physical interactions involve the transfer and conversion of energy, a portion of which is radiated away as sound waves. These sounds are a direct, time-resolved record of the interaction's dynamics.

\begin{itemize}
    \item Contact and Collision Dynamics: The sounds produced by object collisions and friction are direct manifestations of contact mechanics. For brief impacts, Hertzian contact theory models the deformation and forces between two curved surfaces. The duration and peak force of the contact, which directly influence the radiated sound, are functions of the objects' materials (Young's modulus, Poisson's ratio), geometries, and impact velocity. The sound is not only generated by the object's subsequent vibration but also by the rapid acceleration of the air at the contact point. The resulting sound's spectral energy distribution, peak amplitude, and duration are closely related to the mass, relative velocity, hardness, and coefficient of restitution of the colliding bodies \cite{chang2020robot}. If the impact energy exceeds the elastic limit, plastic deformation and fracture can occur, generating distinct high-frequency acoustic emissions. This allows models to not only classify the "falling" event but also regressively estimate the drop height, object mass, and even diagnose damage \cite{fares2023leak, wang2024estimating}. In robotic manipulation, analyzing subtle contact sounds can reveal the stability of a grasp, detect incipient slip, and characterize surface texture, enabling more refined force control and dexterous manipulation \cite{lee2023road, randall2021vibration}.
    
    \item Fluid and Gas Dynamics (Aeroacoustics): The sounds of liquid flow and aerodynamic noise contain rich information about fluid dynamics. The foundational theory is Lighthill's acoustic analogy, which recasts the Navier-Stokes equations into an inhomogeneous wave equation \cite{lighthill1952sound}:
    \begin{equation}
        \frac{\partial^2 \rho'}{\partial t^2} - c_0^2 \nabla^2 \rho' = \frac{\partial^2 T_{ij}}{\partial x_i \partial x_j}
    \end{equation}
    where $\rho'$ is the density fluctuation, $c_0$ is the sound speed in the stationary fluid, and $T_{ij}$ is the Lighthill stress tensor. This powerful equation shows that fluid flow can be modeled as a source of sound (quadrupole in nature) within a stationary acoustic medium. A key extension is the Ffowcs Williams-Hawkings (FW-H) equation, which explicitly includes terms for sound generated by moving surfaces (monopole and dipole sources), making it essential for modeling noise from rotating blades or vehicles. A world model can leverage these principles to monitor pipeline flow regimes \cite{shao2017novel}, detect gas or water leaks from the high-frequency turbulence they generate \cite{cen2022review}, or estimate the fill level of a container by analyzing the changing resonant frequencies of the sloshing liquid \cite{kuttruff2024room}. For autonomous vehicles, the noise generated by tire-road interaction provides real-time information on road conditions (e.g., dry, wet, icy) \cite{song2018deep, brandstein2001microphone}, offering key safety information that is often ambiguous for vision or LiDAR systems.
    
    \item Cyclostationary Analysis for Machine Health: Rotating machinery produces sounds that are not strictly stationary but are cyclostationary—their statistical properties vary periodically with the machine's cycle. When faults like bearing wear or gear tooth cracks develop, they introduce periodic impulsive components or amplitude/frequency modulations into the acoustic signal. Advanced signal processing techniques like spectral correlation and envelope analysis can reveal these hidden periodicities, which are often buried in noise. Deep learning models can then be trained on these processed representations to perform highly sensitive fault diagnosis and prognostics, effectively creating a detailed acoustic physical model of a machine's health state \cite{desai2022review, evers2018acoustic, jiang2019survey}.
\end{itemize}

\subsection{Mapping Spatial Geometry by Acoustic Cues}
The propagation, reflection, diffraction, and reverberation of sound waves within a space create a complex sound field that acts as a sonic fingerprint of the environment's geometry. By analyzing this sound field, an AI can construct a three-dimensional cognitive map of its surroundings.

\begin{itemize}
    \item Room Impulse Response (RIR) as a Geometric Descriptor: The acoustic properties of an enclosed space are comprehensively described by its \textbf{Room Impulse Response (RIR)}, $h(t)$. The signal received at a microphone, $y(t)$, is the convolution of the source signal $s(t)$ with the RIR:
    \begin{equation}
        y(t) = (s * h)(t) = \int_{-\infty}^{\infty} s(\tau) h(t - \tau) d\tau
    \end{equation}
    
    The RIR's structure contains rich geometric information. The early part consists of the direct path sound followed by a series of distinct specular reflections from nearby surfaces. The arrival times and amplitudes of these reflections encode the path lengths and reflective properties of the surfaces, effectively creating an "echogram" of the local geometry. The later part of the RIR, the late reverberation tail, is a diffuse field of countless overlapping echoes. Its decay rate, quantified by the \textbf{Reverberation Time ($T_{60}$)}, is related to the room's volume $V$ and total surface absorption.
    \begin{align}
    T_{60}^{\text{Sabine}} &= \frac{0.161\,V}{A}, 
    &\text{with}\; A=\sum_{i}\alpha_i S_i 
    \quad (\bar{\alpha}\lesssim 0.2) , \label{eq:Sabine} \\
    T_{60}^{\text{Eyring}} &= \frac{0.161\,V}{-S\ln(1-\bar{\alpha})}, 
    &\text{where}\; \bar{\alpha}=A/S 
    \quad (\bar{\alpha}\gtrsim 0.2) . \label{eq:Eyring}
    \end{align}
    Here $S=\sum_i S_i$ is the total surface area, $S_i$ and $\alpha_i$ are the area and absorption coefficient of the $i$-th surface, respectively, and $\bar{\alpha}$ is the average absorption coefficient. Equation~\eqref{eq:Sabine} (Sabine) is accurate for rooms with low average absorption ($\bar{\alpha}\lesssim0.2$), whereas Equation~\eqref{eq:Eyring} (Eyring) remains reliable when absorption is moderate to high ($\bar{\alpha}\gtrsim0.2$) \cite{kuttruff2024room}.

    By analyzing a received sound signal and performing blind deconvolution to estimate the RIR, a model can infer the room's volume, approximate dimensions, and even the types of materials on its surfaces (e.g., reflective glass vs.\ absorptive carpet) \cite{dokmanic2013acoustic, steckel2013batslam}.

    \item Binaural Acoustics and Spatial Hearing: For an agent with two microphones (ears), the brain (or an AI) exploits binaural cues to localize sound sources with remarkable precision. The key cues are:
        \begin{itemize}
            \item \textbf{Interaural Time Difference (ITD):} The difference in arrival time of a sound wave at the two ears, which depends on the source's azimuth.
            \item \textbf{Interaural Level Difference (ILD):} The difference in sound intensity, which is caused by the acoustic shadowing effect of the head and is most prominent at high frequencies.
        \end{itemize}
    The complete set of these spatial cues is captured by the \textbf{Head-Related Transfer Function (HRTF)}, which is the RIR from a source to each of the two ears, a function whose processing is mirrored in the neural representation of three-dimensional acoustic space within the human temporal lobe \cite{zhang2015neural}. By learning a model of its own HRTF, an agent can perform highly accurate 3D sound source localization.
    
    \item Acoustic SLAM and Scene Reconstruction: By combining source localization with agent motion, it's possible to perform Acoustic Simultaneous Localization and Mapping (Acoustic SLAM). By moving through an environment and continuously localizing static sound sources (or using known sound sources as beacons to localize itself), the agent can incrementally build a map of both the sound sources and the primary acoustic reflectors (walls, obstacles) \cite{savioja2015overview, schwarz2000system}. Unlike visual SLAM which relies on optical features, Acoustic SLAM builds a map of the environment's geometric and material properties as they pertain to sound, enabling navigation in complete darkness or visually featureless environments where vision-based systems would catastrophically fail \cite{pertila2009acoustic}. This forms a direct pathway from raw sound to a functional spatial model of the world.
\end{itemize}

\subsection{Modeling Nonlinear Sound Wave Interactions}
While powerful, the linear models discussed so far operate under the assumption of small acoustic perturbations. This assumption breaks down in many real-world scenarios involving high-amplitude sound, such as biomedical ultrasound or close-range robotic interaction. In these regimes, nonlinear effects like harmonic generation and waveform distortion become dominant, requiring more advanced physical models \cite{zuwen1999nonlinear}. A complete physical foundation for acoustic world models must therefore also encompass the principles of nonlinear acoustics.

Nonlinear acoustic models extend linear theory by incorporating amplitude-dependent effects for high-intensity fields. When acoustic pressure is large, linear approximations fail to capture the underlying physics \cite{zuwen1999nonlinear}. By performing a perturbation expansion on the compressible Navier-Stokes equations for an irrotational, lossless flow, we can derive the governing nonlinear wave equations. Let the total pressure be \( p = p_0 + \epsilon p_1 + \epsilon^2 p_2 + \cdots \), where \(p_0\) is ambient pressure and \(\epsilon\) is a small perturbation parameter. Retaining terms up to second order, \(\mathcal{O}(\epsilon^2)\), leads to the Westervelt Equation.

The second-order perturbation expansion of the governing equations yields an intermediate form:
\begin{align}
  \frac{\partial^2 p_1}{\partial t^2} 
    \;-\; c^2 \nabla^2 p_1 
    &= -\,\epsilon \,\frac{\beta}{\rho_0\,c^2} \,\frac{\partial^2 (p_1^2)}{\partial t^2},
  \label{eq:westervelt_perturbation}
\end{align}
where \(\rho_0\) is the ambient density, \(c\) is the linear sound speed, and \(\beta = 1 + \tfrac{B}{2A}\) is the coefficient of nonlinearity. Normalizing this expression leads to the canonical Westervelt Equation:
\begin{equation}
  \frac{\partial^2 p}{\partial t^2} 
    \;-\; c^2 \nabla^2 p 
    = \alpha \frac{\partial^2 (p^2)}{\partial t^2},
  \label{eq:westervelt}
\end{equation}
where \(\alpha \propto \frac{\beta}{\rho_0\,c^2}\) encapsulates the medium’s nonlinear response. This equation models key nonlinear phenomena such as second-harmonic generation and waveform steepening, which are  in high-amplitude acoustic fields.

For directional acoustic beams, the Khokhlov-Zabolotskaya-Kuznetsov (KZK) equation models the combined effects of nonlinearity, diffraction, and absorption. It is derived using a paraxial approximation and a retarded time frame \(\tau = t - z/c\). The general form is:
\begin{equation}
  \frac{\partial^2 p}{\partial z\,\partial \tau}
    =  \frac{c}{2}\,\nabla_\perp^2 p \;+\;  \frac{\beta_{\text{NL}}}{2\,\rho_0\,c^3}\,\frac{\partial^3 (p^2)}{\partial \tau^3}
      \;+\; \frac{\delta}{2\,c^3}\,\frac{\partial^3 p}{\partial \tau^3},
  \label{eq:kzk_full}
\end{equation}
where \(\nabla_\perp^2\) is the transverse Laplacian, \(\beta_{\text{NL}}\) is the nonlinearity coefficient, and \(\delta\) models thermoviscous absorption. Neglecting absorption (\(\delta = 0\)) simplifies the equation. For an axisymmetric beam in cylindrical coordinates, the lossless KZK equation becomes:
\begin{equation}
  \frac{\partial^2 p}{\partial z\,\partial \tau}
    = \frac{c}{2r}\,\frac{\partial}{\partial r}\!\left(r\,\frac{\partial p}{\partial r}\right)
      \;+\; \frac{\beta_{\text{NL}}}{2\,\rho_0\,c^3}\,\frac{\partial^3 (p^2)}{\partial \tau^3}.
  \label{eq:kzk}
\end{equation}
The KZK equation is fundamental for modeling phenomena like focused acoustic holography and other far-field, high-intensity applications \cite{chen2025synergistic}.

\section{Methodological Frameworks: Acoustic World Models}

The inclusion of these nonlinear models highlights a key challenge: as physical models grow in fidelity, their analytical or numerical solutions become intractable, especially for the inverse problems central to building world models. This complexity necessitates a new computational paradigm.

This is precisely the gap that modern AI methods are poised to fill. The core task is to develop a framework that can learn, represent, and reason with these intricate laws of acoustic physics—a problem of system identification and forward simulation. This requires an integration of data-driven machine learning and principle-driven physical modeling. The following section surveys the three key methodological pillars that form this foundation: physics-informed neural networks, generative forward models, and self-supervised multimodal learning.

\subsection{Physics-Informed Neural Networks for Acoustic Inverse Problems}
Physics-Informed Neural Networks (PINNs) represent a fundamental shift in scientific machine learning, moving beyond purely data-driven "black-box" models to create "gray-box" models that are constrained by known physical laws \cite{raissi2019physics, karniadakis2021physics}. A PINN is a neural network that is trained to not only fit observed data but also to satisfy a governing Partial Differential Equation (PDE) over the entire spatiotemporal domain of interest.

\subsubsection{The Core PINN Formulation for Acoustics}
In the context of acoustics, a PINN is typically a Multi-Layer Perceptron (MLP), $\hat{p}(x, t; \theta)$, that takes spatial coordinates $x$ and time $t$ as input and outputs the predicted acoustic pressure $\hat{p}$. The network's parameters $\theta$ are optimized by minimizing a composite loss function:
\begin{equation}
    \mathcal{L}(\theta) = w_{data}\mathcal{L}_{\text{data}} + w_{phys}\mathcal{L}_{\text{phys}} + w_{bc}\mathcal{L}_{\text{bc}} + w_{ic}\mathcal{L}_{\text{ic}}
\end{equation}
where the terms are:
\begin{enumerate}[label=(\roman*)]
    \item Data Fidelity Loss ($\mathcal{L}_{\text{data}}$): This is a standard supervised loss that anchors the solution to real-world measurements. Given a set of $N_{data}$ sensor measurements at points $(x_i, t_i)$, it is typically the Mean Squared Error (MSE):
    $\mathcal{L}_{\text{data}} = \frac{1}{N_{data}} \sum_{i=1}^{N_{data}} |\hat{p}(x_i, t_i; \theta) - p_{\text{measured}}(x_i, t_i)|^2$.
    
    \item Physics Residual Loss ($\mathcal{L}_{\text{phys}}$): This is the heart of the PINN. It enforces the governing PDE, such as the acoustic wave equation, which, for a homogeneous and lossless free sound field, is expressed as the residual function $f(p) = \nabla^2 p - \frac{1}{c^2} \frac{\partial^2 p}{\partial t^2} = 0$. The loss is the MSE of this residual, evaluated at a large number of $N_{\text{phys}}$ collocation points randomly sampled throughout the domain interior:
    $\mathcal{L}_{\text{phys}} = \frac{1}{N_{phys}} \sum_{j=1}^{N_{phys}} |f(\hat{p}(x_j, t_j; \theta))|^2$. The important enabling technology here is automatic differentiation, which allows for the exact computation of the partial derivatives ($\frac{\partial^2 \hat{p}}{\partial t^2}$, etc.) with respect to the network's inputs, directly within the deep learning framework.
    
    \item Boundary and Initial Condition Losses ($\mathcal{L}_{\text{bc}}, \mathcal{L}_{\text{ic}}$): These terms enforce the problem's boundary conditions (e.g., acoustically rigid walls, pressure-release surfaces) and initial state (e.g., pressure and velocity at $t=0$). For example, a Neumann boundary condition $\nabla \hat{p} \cdot \mathbf{n} = 0$ is enforced via an additional MSE term evaluated on the boundary.

\end{enumerate}

The weights $w_{data}, w_{phys}, \dots$ are hyperparameters that balance the different loss components, and their tuning is a important and non-trivial aspect of training PINNs.

\subsubsection{Advantages, Limitations, and Comparative Analysis}
The PINN framework offers profound advantages for acoustic modeling, particularly in data efficiency and solving inverse problems. By embedding physical constraints, PINNs act as a powerful regularizer, transforming ill-posed inverse problems (e.g., source localization from sparse sensors or for volumetric field reconstruction \cite{zhang2024physics, lee2019study, olivieri2024physics}) into well-posed ones. The physics loss provides "free" supervision, reducing the need for extensive labeled data and enabling mesh-free learning.

However, PINNs are not without limitations. Their performance hinges on the availability of an accurate, analytically expressible PDE, which may not exist for highly complex, non-linear, or multi-physics phenomena. They can also struggle with sharp gradients or discontinuities in the solution. Compared to generative models, PINNs excel at inverse problems with sparse data but are less suited for open-ended synthesis of novel sounds. In contrast to self-supervised methods, which learn implicit physical rules from data correspondence, PINNs enforce explicit, pre-defined physical laws, offering stronger guarantees at the cost of flexibility.

\subsection{Generative Models for Acoustics Simulation and Synthesis}
A core function of a world model is to act as a simulator, capable of "imagining" or predicting future sensory inputs. This forward prediction is the basis for planning and reasoning. In the acoustic domain, this translates to building models that can generate realistic audio corresponding to physical events.

\subsubsection{Differentiable Physics Simulation with Gradient Descent}
\textbf{Differentiable simulation} is a powerful approach where the entire simulation pipeline is constructed from differentiable operations \cite{hayes2022mugen}. By re-implementing simulation algorithms (e.g., modal synthesis, Finite-Difference Time-Domain (FDTD) methods) in frameworks like PyTorch, gradients can be backpropagated from a loss function (e.g., the difference between simulated and real audio) all the way back to the initial physical parameters. This enables highly efficient, gradient-based optimization for system identification tasks (e.g., inferring a drum's tension from its sound) and creates powerful data factories for training other models.

\subsubsection{Implicit Neural Models for Acoustic Fields}
Inspired by Neural Radiance Fields (NeRFs) \cite{mildenhall2021nerf}, Neural Acoustic Fields (NAFs) learn a continuous, implicit representation of a scene's acoustic response, typically by mapping a 3D coordinate to a Room Impulse Response (RIR) \cite{gao2024audio}. Once trained on sparse RIR measurements, a NAF can render a physically plausible RIR for any arbitrary point in the scene, providing a fully differentiable, compressed representation of the acoustic environment for applications in Virtual Reality (VR)/Augmented Reality (AR) and robotic path planning. The challenge of maintaining physical consistency under transformations is also addressed by related methods in equivariant neural rendering \cite{dupont2020equivariant}.

\subsubsection{Controllable Procedural Sound Synthesis}
This approach uses deep generative models to directly synthesize raw audio waveforms conditioned on high-level physical parameters. Architectures like Denoising Diffusion Probabilistic Models (DDPMs) \cite{kong2020diffwave} are particularly effective. By conditioning the generation process on a vector of physical parameters (e.g., [mass, material, velocity]), these models learn a direct, controllable mapping from a physical latent space to the audio domain, forming the predictive core of an acoustic world model \cite{dwivedi2018algorithms, ismael2021deep}.

\subsubsection{Comparative Analysis of Generative Approaches}
These generative models offer a powerful alternative to PINNs, particularly for forward prediction and synthesis. Unlike PINNs, which are constrained by a known PDE, generative models can learn to simulate complex phenomena directly from data, even when the underlying physics is poorly understood. However, this flexibility comes at a cost: they typically require significantly more data than PINNs and may generate physically implausible results if not properly regularized. Compared to self-supervised methods, which learn representations for discrimination, generative models learn the full data distribution, making them computationally more intensive but capable of synthesis and prediction.

\subsection{Self-Supervised Multimodal Learning from Unlabeled Data}
To scale learning, an AI must learn from raw, unlabeled sensory data. Self-supervised and multimodal learning paradigms provide a powerful framework for this by leveraging the natural co-occurrence of sensory streams as a supervisory signal.

\subsubsection{Audio-Visual Contrastive Learning}
This paradigm, pioneered by seminal works like \cite{arandjelovic2017look, owens2018audio}, leverages the natural co-occurrence of audio and video. By training a model on a contrastive task—pulling representations of corresponding audio-visual pairs together in an embedding space while pushing non-corresponding pairs apart (e.g., in large-scale multimodal models like Contrastive Language-Image Pre-training (CLIP) \cite{radford2021learning} and Flamingo \cite{alayrac2022flamingo})—the model learns powerful, high-level semantic features. These pre-trained encoders can be fine-tuned for diverse downstream tasks like sound source localization \cite{senocak2018learning} and audio-visual source separation \cite{gao2019co}.

\subsubsection{Action-Conditioned Predictive Learning in Latent Space}
Inspired by frameworks like Dreamer \cite{hafner2019dream}, this approach trains an embodied agent in a closed perception-action loop. The agent learns an encoder to compress sensory input into a latent state, a transition model to predict the next latent state based on an action, and a decoder to reconstruct the predicted sensory input. This process forces the agent to build a compact, internal model of the environment's dynamics directly from interaction, enabling sample-efficient planning in latent space.

\subsubsection{Comparative Analysis of Self-Supervised Learning}
Self-supervised learning offers a powerful way to learn from vast amounts of unlabeled data, a key advantage over supervised methods and even some PINN formulations. Unlike PINNs, which require explicit physical equations, these methods learn implicit physical rules through data association. In contrast to generative models that learn the full data distribution for synthesis, self-supervised methods typically learn discriminative representations optimized for downstream tasks. This often makes them more computationally efficient to train and more effective for tasks like classification or retrieval, but they cannot inherently generate new sensory data.

\subsection{Comparative Analysis and Hybrid Approaches}
The three methodological pillars—PINNs, generative models, and self-supervised learning—offer distinct trade-offs, summarized in Table \ref{tab:methods_comparison}. No single approach is universally superior; the optimal choice depends on the specific problem, data availability, and computational constraints.

The future of acoustic world models likely lies in \textbf{hybrid approaches} that combine the strengths of these pillars. For instance, a self-supervised model could be used to pre-train a powerful feature extractor on vast unlabeled video data. A generative model could then be conditioned on these features to synthesize physically plausible sounds. Finally, a PINN-based loss term could be used during the generative model's training to act as a physical regularizer, ensuring the synthesized audio does not violate known acoustic principles. Such a hybrid system would leverage the data-efficiency of PINNs, the expressive power of generative models, and the scalability of self-supervised learning.

\begin{table}[H]
\centering
\caption{Comparative analysis of core methodologies for acoustic world models}
\label{tab:methods_comparison}
\resizebox{\textwidth}{!}{%
\begin{tabular}{>{\raggedright}p{0.15\textwidth} >{\raggedright}p{0.2\textwidth} >{\raggedright}p{0.22\textwidth} >{\raggedright}p{0.2\textwidth} >{\raggedright\arraybackslash}p{0.23\textwidth}}
\toprule
\textbf{Methodology} & \textbf{Data requirement} & \textbf{Computational cost} & \textbf{Physical fidelity} & \textbf{Primary use case} \\
\midrule
\textbf{Physics-informed neural networks} & Low (sparse labeled data + collocation points). & Moderate to high (training involves PDE evaluation). & High (explicitly constrained by physical laws). & Solving inverse problems (e.g., source localization, material identification). \\
\midrule
\textbf{Generative forward models} & High (requires large datasets, often paired with physical parameters). & High (especially for training diffusion models or complex simulators). & Variable (can be high with good data, but may generate implausible results). & Forward simulation, controllable sound synthesis, data augmentation, virtual environments. \\
\midrule
\textbf{Self-supervised multimodal learning} & Very high (unlabeled multimodal data, e.g., video). & High (training large encoders like transformers). & Implicit (learns correlations, not explicit laws).  & Learning robust representations for downstream tasks (classification, retrieval, source separation). \\
\bottomrule
\end{tabular}%
}
\end{table}

\begin{figure}[H]
    \centering
    \includegraphics[width=1.0\textwidth]{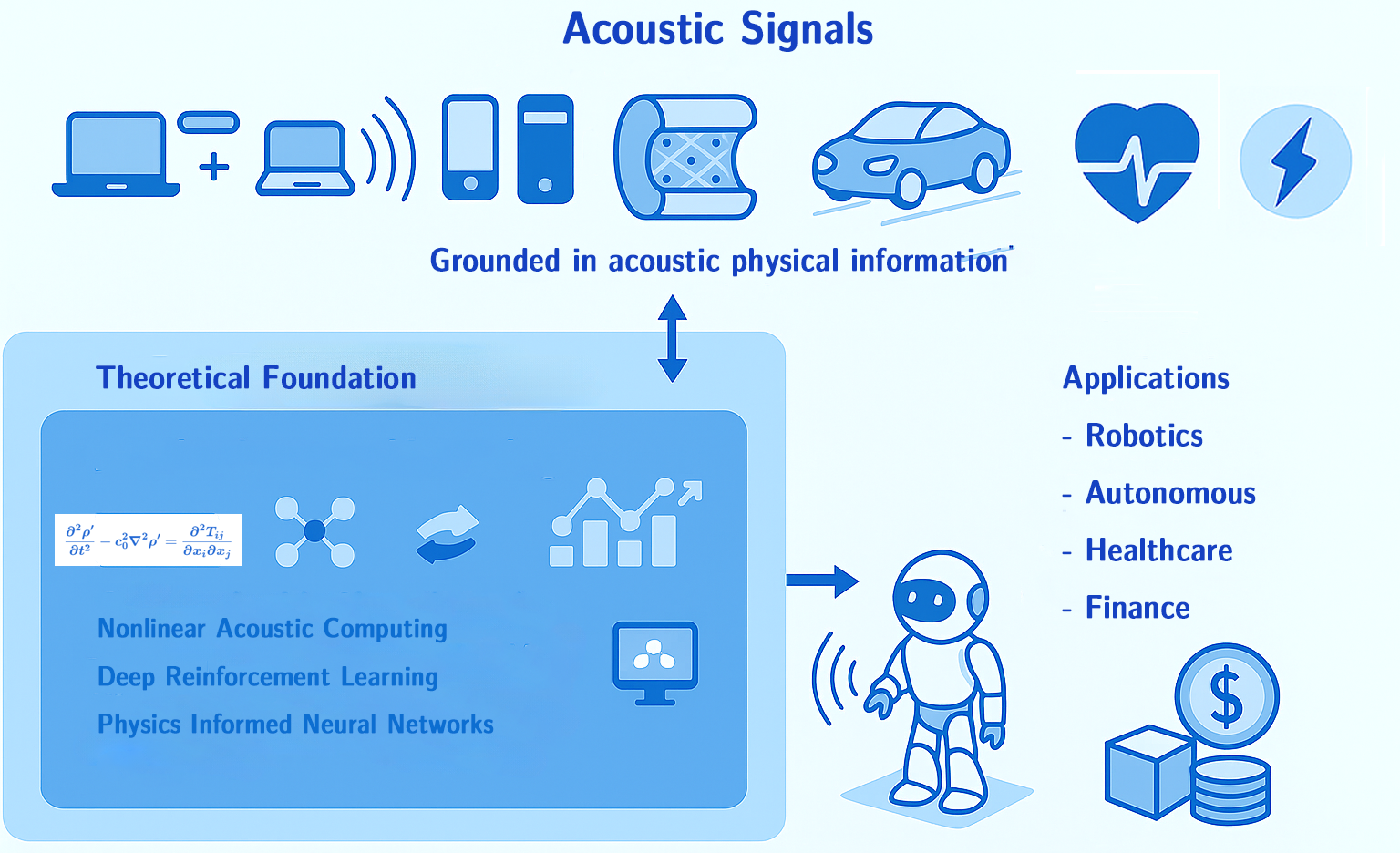}
    \caption{A conceptual diagram illustrating the methodological framework for acoustic world models.}
    \label{fig:methodology}
\end{figure}

\textit{It showcases the interplay between three pillars: (1) Physics-Informed Learning (e.g., PINNs), which embeds physical laws as constraints; (2) Generative Forward Models (e.g., Differentiable Simulators, NAFs), which enable prediction and data synthesis; and (3) Self-Supervised Multimodal Learning, which learns rich representations from unlabeled data. The arrows indicate the flow of information, constraints, and learned representations, forming a cohesive system. With a solid understanding of the theoretical underpinnings and the core computational methodologies, we can now turn our attention to the tangible impact of these models. The next section will demonstrate how acoustic world models are being deployed in real-world systems, highlighting their transformative applications across robotics, autonomous driving, healthcare, and beyond.}

\section{Embodied Applications: Deploying Acoustic World Modes in Real World}

Acoustic world models have the potential to drive significant advancements in various domains of embodied intelligence, where agents must perceive, reason, and act within the complex, dynamic, and often uncertain physical world.

\subsection{Robotics}
In robotics, acoustic perception is not a luxury but a necessity for overcoming the limitations of vision and proprioception, enabling safer, more precise, and more intuitive interaction with the world.
\begin{itemize}
    \item \textbf{Robust Navigation and Mapping in Extreme Environments:} In visually degraded environments (e.g., smoke-filled rooms, complete darkness, dust clouds from drilling), robots can employ Acoustic SLAM for localization and mapping. By emitting active sonar pings or passively listening to environmental sounds, the robot can construct a geometric map of its surroundings where vision-based systems would completely fail, ensuring operational continuity in important scenarios like search-and-rescue or mining \cite{savioja2015overview, pertila2009acoustic}.
    \item \textbf{Dexterous Manipulation through Acoustic Haptics:} By learning to "listen" to contact sounds, a robot can gain a rich, haptic-like understanding of its physical interactions. It can detect the precise moment of contact, estimate contact force from impact sound intensity, perceive surface textures from frictional sounds, and detect high-frequency transients that signal incipient slip during grasping, allowing for real-time grip adjustment \cite{lee2023road, randall2021vibration}. This acoustic feedback loop enables far more stable and delicate manipulation than is possible with low-frequency force sensors alone. Furthermore, robots can infer latent object properties, such as whether a container is full, empty, or contains a liquid, by shaking it and analyzing the resulting sloshing or rattling sounds, a task impossible for vision, building on a long history of behavior-grounded robotics where agents learn object properties through multimodal interaction \cite{sinapov2011object, bonarini2020communication, chang2020robot}.
    \item \textbf{Intuitive Human-Robot Collaboration:} In collaborative settings, acoustic cues are vital for fluid and safe interaction. A robot can understand a human's activity and intent by listening to tool sounds (e.g., distinguishing drilling from sawing to anticipate the next step in an assembly task, a capability that relies on robust polyphonic sound event detection systems \cite{chan2020comprehensive}). More subtly, it can interpret the paralinguistic and prosodic features of human speech (pitch, tone, speaking rate) to infer their cognitive load, confidence, or emotional state, leading to safer, more adaptive, and more socially aware Human-Robot Collaboration (HRC)\cite{chen2025synergistic, mittal2023acoustic}.
\end{itemize}

\subsection{Autonomous Driving}
While vision, LiDAR, and radar are the primary sensors for autonomous vehicles, acoustics provides a important, omnidirectional, and computationally inexpensive complementary channel for enhanced situational awareness and safety.
\begin{itemize}
    \item \textbf{360-Degree Hazard Perception and Safety Enhancement:} Microphone arrays can reliably detect and localize the sirens of emergency vehicles long before they are visible or detectable by directional sensors, allowing the vehicle to yield preemptively \cite{crocco2016audio}. They can also identify potential threats in the vehicle's blind spots, such as a nearby motorcycle engine, a bicycle's bell, or a child's shout, providing an essential layer of safety. Critically, by analyzing tire-road noise, an acoustic world model can classify road surface conditions (dry, wet, icy, gravel) in real-time by recognizing their distinct spectral signatures. This provides vital input for adaptive traction control and braking systems, a task where vision can be ambiguous \cite{song2018deep, lee2023road, brandstein2001microphone}.
    \item \textbf{Vehicle Prognostics and Health Monitoring (VPHM):} An onboard acoustic world model can continuously "listen" to the vehicle's own sounds, monitoring the engine, brakes, suspension, and other electromechanical components. By building a model of the vehicle's normal acoustic signature, it can detect subtle deviations—anomalous squeals, rattles, or hums—that are early indicators of mechanical failure, such as bearing wear or exhaust leaks. This enables predictive maintenance, improving overall vehicle safety and reliability \cite{evers2018acoustic, desai2022review}.
\end{itemize}

\subsection{Healthcare and Personalized Care}
Acoustic world models have the potential to transform ubiquitous devices like smartphones and smart speakers into powerful, non-invasive tools for continuous health monitoring and early disease diagnosis.
\begin{itemize}
    \item \textbf{Remote Digital Biomarkers for Disease Screening:} As previously mentioned, by modeling the acoustic physics of the respiratory system, AI models can analyze the spectral and temporal properties of coughs, breathing, and snoring sounds to screen for a wide range of conditions like COVID-19, asthma, pneumonia, and obstructive sleep apnea, enabling large-scale, low-cost public health initiatives \cite{ismael2021deep, cummins2015review, laguarta2020covid}. The field is expanding to include gastrointestinal acoustics for digestive health and joint acoustics for osteoarthritis monitoring.
    \item \textbf{Neurological and Mental Health Assessment:} The human voice is a rich source of physiological and neurological biomarkers. An acoustic model can quantify subtle changes in vocal features—such as pitch variability (jitter), amplitude variability (shimmer), speech rate, and harmonic-to-noise ratio—to non-invasively monitor the progression of neurological conditions like Parkinson's disease, enabling monitoring that can inform adaptive therapies \cite{di2021adaptive}, and detect mental health states such as depression, anxiety, and PTSD \cite{koolagudi2012emotion, wani2021comprehensive}.
    \item \textbf{Empathetic AI and Socially Assistive Agents:} For AI companions and virtual assistants, understanding the user's emotional state is paramount for effective interaction. By analyzing the emotional prosody of speech—the "music" behind the words—an AI can infer joy, sadness, anger, or stress from pitch contours, intensity, and rhythm. This allows the AI to provide more empathetic, appropriate, and natural responses, fostering deeper human-machine trust and rapport in applications like elderly care and mental health support \cite{baltruvsaitis2018multimodal}.
\end{itemize}

\subsection{Financial Services}
Acoustic data in the financial sector, particularly from earnings calls, client interactions, and trading floors, is rich in emotional, cognitive, and paralinguistic features. The acoustic world models offer a shift from passively "hearing" to actively "understanding" contextual and causal cues.
\begin{itemize}
    \item \textbf{Quantitative Investment and Alpha Generation:} Acoustic world models can extract non-consensus alpha signals from public acoustic sources like quarterly corporate earnings calls. By analyzing subtle paralinguistic cues in executives' voices—such as vocal stress (measured by jitter and shimmer), pitch modulation, hesitation rates, and response latencies—these models can generate quantitative metrics of managerial confidence, deception, or uncertainty. These acoustic biomarkers have been shown to be orthogonal to text-based sentiment analysis and predictive of future firm performance and stock returns \cite{mayew2012power}.
    \item \textbf{Advanced Risk Management and Compliance:} In compliance and risk management, acoustic world models can enhance fraud detection by quantifying acoustic markers of deceptive or high-stress speech patterns in trader communications or client calls. This allows for proactive risk alerts that go beyond simple keyword spotting, providing a deeper layer of security and monitoring for detecting financial misreporting or market manipulation activities \cite{hobson2012analyzing}.
    \item \textbf{Customer Interaction and Knowledge Engineering:} In client-facing roles, acoustic world models can analyze conversational dynamics to optimize service quality. By understanding nuances in tone, turn-taking dynamics, and engagement levels from call center audio, financial institutions can automatically predict customer satisfaction, identify at-risk clients, and provide real-time feedback to agents, thereby improving operational efficiency and client retention \cite{chowdhury2016predicting}.
\end{itemize}

\begin{longtable}{>{\raggedright\arraybackslash}p{0.13\textwidth} >{\raggedright\arraybackslash}p{0.18\textwidth} >{\raggedright\arraybackslash}p{0.18\textwidth} >{\raggedright\arraybackslash}p{0.18\textwidth} >{\raggedright\arraybackslash}p{0.18\textwidth}}
\caption{Summary of applications for acoustic world models}\label{tab:applications}\\
\toprule
\textbf{Application domain} & \textbf{Specific task} & \textbf{Key acoustic information} & \textbf{Enabling technology} & \textbf{Primary impact / advantage}\\
\midrule
\endfirsthead

\multicolumn{5}{c}{\bfseries \tablename\ \thetable{} -- continued from previous page}\\
\toprule
\textbf{Application domain} & \textbf{Specific task} & \textbf{Key acoustic information} & \textbf{Enabling technology} & \textbf{Primary impact / advantage}\\
\midrule
\endhead

\bottomrule
\endfoot

\multirow{3}{*}{\textbf{Robotics}}
& Dexterous manipulation  & High-frequency transients from contact friction & Self-supervised; anomaly detection & Enhanced dexterity and grasp stability; prevention of object damage.\\
\cmidrule{2-5}
& Object property inference & Sloshing/rattling sounds from shaking actions & Action-conditioned predictive models & Perception of latent properties invisible to vision; richer interaction.\\
\cmidrule{2-5}
& Navigation in degraded environments & Echoes from active sonar or passive environmental sounds & Acoustic SLAM, Differentiable Simulation & Robust localization and mapping in darkness, smoke, or dust where vision fails.\\
\midrule
\multirow{3}{*}{\textbf{Autonomous}}
& Emergency vehicle detection & Siren harmonics and Doppler shift & Microphone array processing, beamforming & Enhanced safety through early, 360-degree hazard awareness.\\
\cmidrule{2-5}
& Road surface classification & Spectral signatures of tire-road noise & Supervised learning & Improved safety via adaptive traction control and braking systems.\\
\cmidrule{2-5}
& Vehicle health monitoring & Anomalous squeals, rattles, or hums from components & Anomaly detection, cyclostationary analysis & Predictive maintenance, increased vehicle reliability and safety.\\
\midrule
\multirow{3}{*}{\textbf{Healthcare}}
& Respiratory disease screening & Spectral and temporal features of coughs/breathing & Deep learning classifiers, generative models & Low-cost, non-invasive, and scalable public health screening.\\
\cmidrule{2-5}
& Neurological health assessment & Vocal biomarkers (jitter, shimmer, prosody) & Regression and classification models on speech features & Non-invasive tracking of disease progression and mental state.\\
\cmidrule{2-5}
& Empathetic AI companions & Emotional prosody in speech (pitch, tone, rhythm) & Multimodal emotion recognition & More natural, effective, and supportive human-machine interaction.\\
\midrule
\multirow{3}{*}{\textbf{Financial}}
& Alpha generation from earnings calls & Paralinguistic cues of stress/confidence in executive voices & Regression models on vocal features & Orthogonal investment signals not found in text; enhanced prediction.\\
\cmidrule{2-5}
& Fraud detection and compliance & Acoustic markers of deceptive or high-stress speech & Anomaly detection, behavioral modeling & Proactive risk management and enhanced security beyond keyword spotting.\\
\cmidrule{2-5}
& Customer service optimization & Conversational dynamics, turn-taking, vocal engagement & Emotion recognition, conversational AI & Improved client satisfaction, retention, and operational efficiency.\\
\end{longtable}

\textit{The diverse applications showcased in this section underscore the immense practical value of acoustic intelligence. However, the path to widespread, robust, and ethical deployment is not without significant hurdles. Therefore, the final section of this survey provides a important examination of these challenges and outlines a forward-looking research agenda to guide the field's future development.}

\section{Challenges, Ethics, and Future Outlook}

Despite the immense potential of acoustic world models, their transition from laboratory concepts to robust, real-world deployments faces significant scientific challenges and serious ethical considerations. A clear-eyed assessment of these hurdles is essential for guiding future research and ensuring responsible innovation. This section first details the pressing technical and societal challenges, then synthesizes the state of the art to propose a detailed, multi-faceted research agenda for the next generation of acoustic intelligence.

\subsection{Broad Scientific and Technical Challenges}
The core scientific challenges lie at the intersection of high-dimensional data processing, physical modeling, and resource-constrained computation.

\begin{itemize}
    \item \textbf{High-Dimensional and Asynchronous Multimodal Fusion:} The physical world provides a symphony of sensory data. Effectively fusing heterogeneous, asynchronous, and often noisy data streams from acoustics, vision, haptics, and other sensors into a cohesive representation remains a fundamental research problem \cite{baltruvsaitis2018multimodal}. The challenge is not merely concatenation. It involves learning the complex, non-linear correlations and causal dependencies between modalities. For example, the high-frequency transient of an impact sound occurs before the object's visual motion ceases. Architectures must move beyond simple early, late, or hybrid fusion schemes towards models that can explicitly handle temporal asynchrony and learn a shared latent space where one modality can inform or correct another. Techniques like multimodal transformers with cross-attention mechanisms are promising, but efficiently scaling them and learning their alignment without massive supervision is an open question.

    \item \textbf{Computational Complexity and Real-Time Edge Deployment:} A stark trade-off exists between physical fidelity and computational feasibility. High-fidelity acoustic simulations based on wave-based methods (e.g., FDTD, Boundary Element Method (BEM)) are computationally prohibitive for real-time applications, often requiring hours of computation on high-performance clusters for a few seconds of audio. Conversely, large deep learning models like diffusion models or large transformers, while powerful, are difficult to deploy on resource-constrained edge devices such as robots, wearables, or vehicles. This necessitates a multi-pronged research effort into: (a) \textbf{Model Compression}, including techniques like network pruning, structured weight matrices, trained quantization, and Huffman coding \cite{han2015deep}; (b) \textbf{Knowledge Distillation}, where a large, complex "teacher" model (e.g., a full physics simulator) is used to train a smaller, faster "student" model (a neural network); and (c) \textbf{Efficient Architectures}, exploring alternatives like Spiking Neural Networks (SNNs) or state-space models that are inherently more computationally and power-efficient;  and (d) \textbf{Hardware-Software Co-Design}, recognizing that model-centric optimizations are only one side of the equation. The co-design of learning algorithms with specialized, power-efficient silicon—such as Neural Processing Units (NPUs), Tensor Processing Units (TPUs), or custom ASICs—is paramount for achieving real-time inference on the edge. This synergy is particularly important when leveraging advanced sensors like high-density MEMS (Micro-Electro-Mechanical Systems) microphone arrays or highly sensitive fiber-optic acoustic sensors. While these sensors provide richer, higher-fidelity input data that enhances perceptual capabilities, they also intensify the data throughput and computational demand, demanding a holistic, system-level approach that tightly integrates sensor technology, hardware acceleration, and algorithmic efficiency.

    \item \textbf{The Acoustic Sim-to-Real-to-Sim Gap:} A major hurdle in applying models trained on synthetic data is the discrepancy between clean, idealized simulated acoustics and noisy, complex, real-world recordings. The "sim-to-real" gap arises from unmodeled phenomena: ambient noise, non-linear material properties, complex reverberation patterns, and the specific frequency response of the microphone and data acquisition hardware. Bridging this gap requires advanced techniques like: (a) \textbf{Domain Randomization}, where simulation parameters (noise levels, reverberation times, material properties) are heavily randomized during training to force the model to learn robust, invariant features; (b) \textbf{System Identification}, where a short calibration phase is used to learn a model of the specific sensor's transfer function, which can then be applied to the simulator's output; and (c) \textbf{Unsupervised Domain Adaptation}, which attempts to align the feature distributions of the simulated and real domains without requiring paired data \cite{wang2018supervised}. A further challenge is the "real-to-sim" problem: accurately capturing a real-world scene's geometry and material properties to create a high-fidelity digital twin for simulation in the first place.
\end{itemize}

\subsection{Ethics and Regulation of Acoustic World Models}
The power of acoustic sensing brings with it a corresponding responsibility. Failure to proactively address the ethical dimensions of this technology could lead to significant societal harm. In what follows, we examine four principal ethical pressure-points and articulate concrete governance mechanisms. For clarity and academic rigor, each item is structured into three cohesive components: (i) a concise thematic statement, (ii) an illustrative \textit{Case Study}, and (iii) a detailed discussion of \textit{Response Strategies}. This tripartite format not only sharpens the analytical lens but also furnishes practitioners with actionable guidance.

\begin{itemize}
    \item \textbf{Privacy, Monitoring, and Safety in Sound Space:} 
    The “always-on” nature of pervasive microphones transforms every domestic or public venue into a potential data-collection node. Without guardrails, such ubiquity risks normalizing involuntary surveillance and eroding long-standing social expectations regarding private speech. \par
    
    \textit{Case Study.} A smart-home device designed to detect a ``glass-break'' sound for security must inevitably process all ambient sounds. This includes the inadvertent capture of sensitive family conversations, private health discussions, or interpersonal disputes. If such raw data are stored or transmitted, they could be repurposed by third parties—for instance, to infer purchasing preferences, target advertising, or even facilitate state-level surveillance—well beyond the scope originally consented to. \par
    
    \textit{Response Strategies.} A multi-layered, \emph{privacy-by-design} architecture is essential. (a) \textbf{Mandatory On-Device Processing}: deploy specialized low-power hardware—e.g., Neural Processing Units (NPUs) or Tensor Processing Units (TPUs)—so that raw audio never exits the device. (b) \textbf{Event-Triggered Data Handling}: maintain a low-power, non-recording state until a local acoustic signature (``wake word'' or event) is detected. (c) \textbf{Federated Learning}: improve models on-device, uploading only anonymized, aggregated parameter updates to a central server \cite{konevcny2016federated}. (d) \textbf{Differential Privacy}: employ formal guarantees ensuring that analysis outputs remain stable even if any single individual’s data are removed, thus preventing re-identification \cite{yang2024privacy}. \par

    \item \textbf{Algorithmic Bias, Fairness, and Acoustic Equity:} 
    An AI model naturally mirrors the statistical distribution of its training data; when that data are unbalanced, inequitable outcomes can ensue. In health care, such biases can directly translate into life-altering disparities. \par
    
    \textit{Case Study.} Consider an AI diagnostic tool trained to detect early signs of Parkinson’s disease from vocal tremors. If the training corpus consists predominantly of male speakers of North-American English, performance may degrade for female speakers, individuals with strong regional accents, or non-native speakers. The resultant misdiagnoses delay treatment for under-represented groups and exacerbate existing health inequities. \par
    
    \textit{Response Strategies.} A proactive, end-to-end strategy is required: (a) \textbf{Diverse and Representative Data Curation} that explicitly targets demographic balance (age, gender, ethnicity, accent). (b) \textbf{Bias Auditing} prior to deployment, employing fairness metrics such as equalized odds or demographic parity to assess subgroup performance. (c) \textbf{Fairness-Aware Algorithms}, e.g., adversarial debiasing or re-weighting loss functions to penalize performance disparities \cite{jia2018transfer}. \par

    \item \textbf{Accountability, Transparency, and Explainable AI:}
    In safety-critical domains, opaque “black-box” outputs are ethically indefensible. Stakeholders—from system engineers to end-users—require intelligible justifications for model behavior in both nominal and failure modes. \par
    
    \textit{Case Study.} An autonomous vehicle’s acoustic subsystem misclassifies an icy road patch as merely wet, failing to engage traction-control protocols and leading to an accident. Post-incident analysis hinges on understanding \emph{why} the model erred: was a salient frequency component misinterpreted, or did domain shift cause over-confidence? \par
    
    \textit{Response Strategies.} Explainable AI tools for acoustic signals are indispensable: (a) \textbf{Saliency Maps and Feature Attribution}—e.g., Gradient-weighted Class Activation Mapping (Grad-CAM) on spectrogram inputs—to highlight influential time-frequency bins. (b) \textbf{Counterfactual Explanations} that articulate, in human-centric terms, how slight spectral perturbations would have altered the outcome (``Had the 2\,kHz peak been 5\,dB lower, the surface would have been labeled ‘icy.’''). (c) \textbf{Model-Level Explanations} offering global rule abstractions to expose systematic flaws rather than isolated misfires \cite{samek2021explaining}. \par

    \item \textbf{Risk of Exploitation and Harmful Use:}
    Technologies that sense affective vocal markers confer both empathetic benefits and coercive power. Unchecked, they may be weaponized in disinformation, exploitative marketing, or asymmetric warfare. \par
    
    \textit{Case Study.} Vocal stress-detection algorithms—originally intended for fraud prevention in finance \cite{mayew2012power, hobson2012analyzing}—could be re-deployed in coercive interrogations or micro-targeted political messaging, nudging individuals at their most vulnerable moments. \par
    
    \textit{Response Strategies.} To neutralize dual-use threats, we advocate a trident approach: (a) \textbf{Robust Ethical Frameworks} codified in international standards; (b) \textbf{Regulatory Oversight} with mandatory auditing of high-risk applications; and (c) \textbf{Deliberative Public Discourse} that surfaces societal values and informs policy. \par
\end{itemize}

\subsection{Future Research Agenda for the Next Decade}
Looking ahead, the field must move beyond descriptive pattern recognition towards a generative and causal understanding of the physical world through sound. We outline five interconnected, high-impact research pathways that will define the next generation of acoustic world models.

\subsubsection{Pathway I: From Correlation to Causality — Building Causal Acoustic World Models}
Current models excel at learning correlations. The next frontier is learning the underlying causal relationships to answer counterfactual and interventional questions.

\begin{itemize}
    \item \textbf{Integration with Structural Causal Models (SCMs):}
    The goal is to represent the world as a causal graph, thereby empowering acoustic systems to reason about \emph{why} events occur and to predict the outcome of hypothetical interventions. Such capability marks a decisive shift from passive observation to active, question-driven inference.

    \par\textit{Technical Difficulty:} The search space for causal graph structures is combinatorially explosive, and structures are often non-identifiable from purely observational data. Confronting these obstacles demands an innovation that collapses experimental effort while simultaneously expanding causal identifiability.

    \par\textit{Breakthrough Point:} Developing algorithms that can leverage minimal, targeted interventions (e.g., a robot actively tapping an object) to drastically prune the space of possible causal graphs.

    \item \textbf{Counterfactual Sound Synthesis:}
    Building generative models that can simulate interventions via the ``do-operator'' would allow researchers to ask, and answer, what-if questions about acoustic phenomena. Such models would form the basis of virtual laboratories for sound. To reach that goal, the field must overcome a set of fundamental representation-learning bottlenecks.

    \par\textit{Technical Difficulty:} This requires perfect disentanglement of the causal factors of sound generation (e.g., shape, material, action), which current generative models struggle with. Addressing this challenge will hinge on unifying disentangled latent learning with physically grounded generation.

    \par\textit{Breakthrough Point:} Combining Variational Autoencoders (VAEs) for disentanglement with physics-constrained generative models (e.g., PINN-regularized Generative Adversarial Networks (GANs)) that enforce the plausibility of counterfactual outcomes.
\end{itemize}

\subsubsection{Pathway II: From Determinism to Distributions — Probabilistic and Uncertainty-Aware Models}
The next generation of world models must be probabilistic, predicting a full distribution over possible future sounds and quantifying their own uncertainty.

\begin{itemize}
    \item \textbf{Distinguishing Uncertainty Types:}
    Future acoustic agents must clearly differentiate between randomness in the world and ignorance in the model. Such nuance is important for risk-aware decision-making and for allocating computational resources to learning where it matters most. A principled decomposition, however, introduces non-trivial architecture and training-time complications.

    \par\textit{Technical Difficulty:} Decomposing these two sources of uncertainty within a single, efficient deep learning model is a standing challenge. Any viable solution must not only separate these uncertainties but also exploit them to drive safer interaction policies.

    \par\textit{Breakthrough Point:} Creating hybrid architectures, for instance, using Deep Ensembles to capture uncertainty while having each ensemble member output a full probabilistic distribution (e.g., via a mixture density network) to capture aleatoric uncertainty.

    \item \textbf{Uncertainty-Informed Decision Making:}
    An intelligent agent should convert calibrated uncertainty into concrete behavioral choices—slowing down, gathering more data, or selecting conservative actions—thereby achieving a measurable reduction in risk. Translating scalar uncertainty into high-dimensional control signals, however, is far from straightforward.

    \par\textit{Technical Difficulty:} Translating a raw uncertainty value into a concrete policy change (e.g., how much to slow down a robot arm) is non-trivial. Overcoming this barrier will require algorithms that couple prediction fidelity with action optimization in a single computational loop.

    \par\textit{Breakthrough Point:} Tightly integrating uncertainty quantification with model-based Reinforcement Learning (RL), where the planning algorithm explicitly seeks to minimize worst-case outcomes over the predicted distribution or chooses actions that reduce epistemic uncertainty.
\end{itemize}

\subsubsection{Pathway III: From Single Modes to Unified Perception — Physics-Based Cross-Modal Generation}
The ultimate goal is a model that can ``imagine'' across modalities in a physically consistent way (e.g., generate sound from a silent video and vice versa).

\begin{itemize}
    \item \textbf{A ``Multimodal Turing Test'' for Physics:}
    A rigorous benchmark should require fine-grained, physically plausible cross-modal generation—capturing not only appearance and timing but also material properties and energy dissipation.  Delivering on this benchmark will necessitate an unprecedented synergy of generative learning and explicit physics priors.

    \par\textit{Technical Difficulty:} Ensuring precise temporal synchronization (e.g., sound of footstep at the exact frame of impact) and physical consistency (e.g., sound decay matches visual material properties) is extremely difficult. The key lies in embedding differentiable physics directly into the adversarial training loop, thus allowing the discriminator to penalize implausible patterns at train-time.

    \par\textit{Breakthrough Point:} Developing physics-constrained adversarial training loops, where a discriminator network is trained to judge not only the realism of a generated modality but also its consistency with physical laws, potentially guided by a differentiable physics engine as a third ``umpire.''

    \item \textbf{Unified Latent Space Architectures:}
    Achieving modality-invariant representations will enable zero-shot transfer and few-shot learning across diverse sensory inputs. Yet building a shared latent manifold introduces its own representational pathologies.

    \par\textit{Technical Difficulty:} Naively sharing a latent space often leads to ``modality collapse,'' where one modality dominates the representation. A carefully structured latent hierarchy could offer a pathway to balanced, interpretable multimodal abstraction.

    \par\textit{Breakthrough Point:} Designing architectures with structured, factorized latent spaces and specialized cross-modal attention mechanisms that enforce a ``common language'' for abstract physical concepts (e.g., `rigidity`, `impact\_force`) rather than just low-level features.
\end{itemize}

\subsubsection{Pathway IV: From Passive Hearing to Active Perception — Physical and Exploratory Agents}
The future lies with embodied agents that are active sensors, performing experiments to learn about the world, inspired by echolocating animals \cite{steckel2013batslam}.

\begin{itemize}
    \item \textbf{Information-Theoretic Action Selection:}
    An optimal agent strategically selects actions that maximize information gain, thereby turning the world into an adaptive curriculum for self-supervised learning. Unfortunately, the required information-gain computations are often intractable in continuous action spaces.

    \par\textit{Technical Difficulty:} Calculating the expected information gain for all possible actions in continuous action spaces is often intractable.  Overcoming this bottleneck will involve new approximation paradigms that scale with dimensionality while preserving theoretical guarantees.

    \par\textit{Breakthrough Point:} Developing efficient approximation methods, such as learning a ``value of information'' function that directly predicts the utility of an action for model improvement, turning the information-gathering problem into a standard RL problem.

    \item \textbf{The Perception-Action Loop Dilemma:}
    Building a refined world model and discovering informative actions are tightly coupled tasks, each dependent on the other for progress. Breaking this cycle requires mechanisms for staged acquisition of competence.

    \par\textit{Technical Difficulty:} This is a ``chicken-and-egg'' problem: a good world model is needed to select informative actions, but informative actions are needed to build a good model.  Curriculum strategies that reward learning progress, rather than immediate task success, offer a promising escape from this loop.

    \par\textit{Breakthrough Point:} Employing curriculum learning and intrinsic motivation. The agent starts with simple, random ``babbling'' actions to build a crude initial model, then is rewarded for actions that lead to the largest improvement (i.e., highest prediction error) in its world model, driving it to explore the boundaries of its knowledge.
\end{itemize}

\subsubsection{Pathway V: 	From Custom Models to General AI —  Core Model Framework for Acoustic Physics}
The grand challenge is to train a massive, multi-task, multimodal foundation model on the entirety of available acoustic-physical data.

\begin{itemize}
    \item \textbf{Unprecedented Data Scale and Heterogeneity:}
    Curating petabyte-scale corpora across simulation, robotics, and real-world recordings will supply the raw fuel required to train a genuinely world model. The logistical and engineering demands of such a corpus, however, are without historical precedent.

    \par\textit{Technical Difficulty:} Harmonizing these diverse data sources with varying labels, noise levels, and formats is a significant engineering challenge. Real progress will require pre-training objectives that thrive on noise and exploit complementary structure across modalities.

    \par\textit{Breakthrough Point:} Developing sophisticated self-supervised pre-training objectives that are robust to noise and can learn from multiple data modalities simultaneously, such as a physics-informed version of masked autoencoding across audio, video, and text tokens.

    \item \textbf{Architectural and Computational Scalability:}
    Scaling up from niche prototypes to a true foundation model demands both algorithmic ingenuity and hardware-software co-design tailored to continuous signals. The resource footprint of a naive transformer approach exceeds the budgets of even well-funded laboratories.

    \par\textit{Technical Difficulty:} The computational cost is extremely high, and standard transformer architectures may not be optimal for continuous, physical signals. Progress will hinge on a class of models whose capacity grows sub-linearly with compute while maintaining expressive power.

    \par\textit{Breakthrough Point:} Innovations in efficient model architectures (e.g., state-space models, mixture-of-experts) and specialized hardware/software co-design to make training such a model feasible. The ultimate vision is a single model that can be fine-tuned for a vast range of physical acoustic tasks with minimal new data.
\end{itemize}

\begin{figure}[H]
    \centering
    \includegraphics[width=1.0\textwidth]{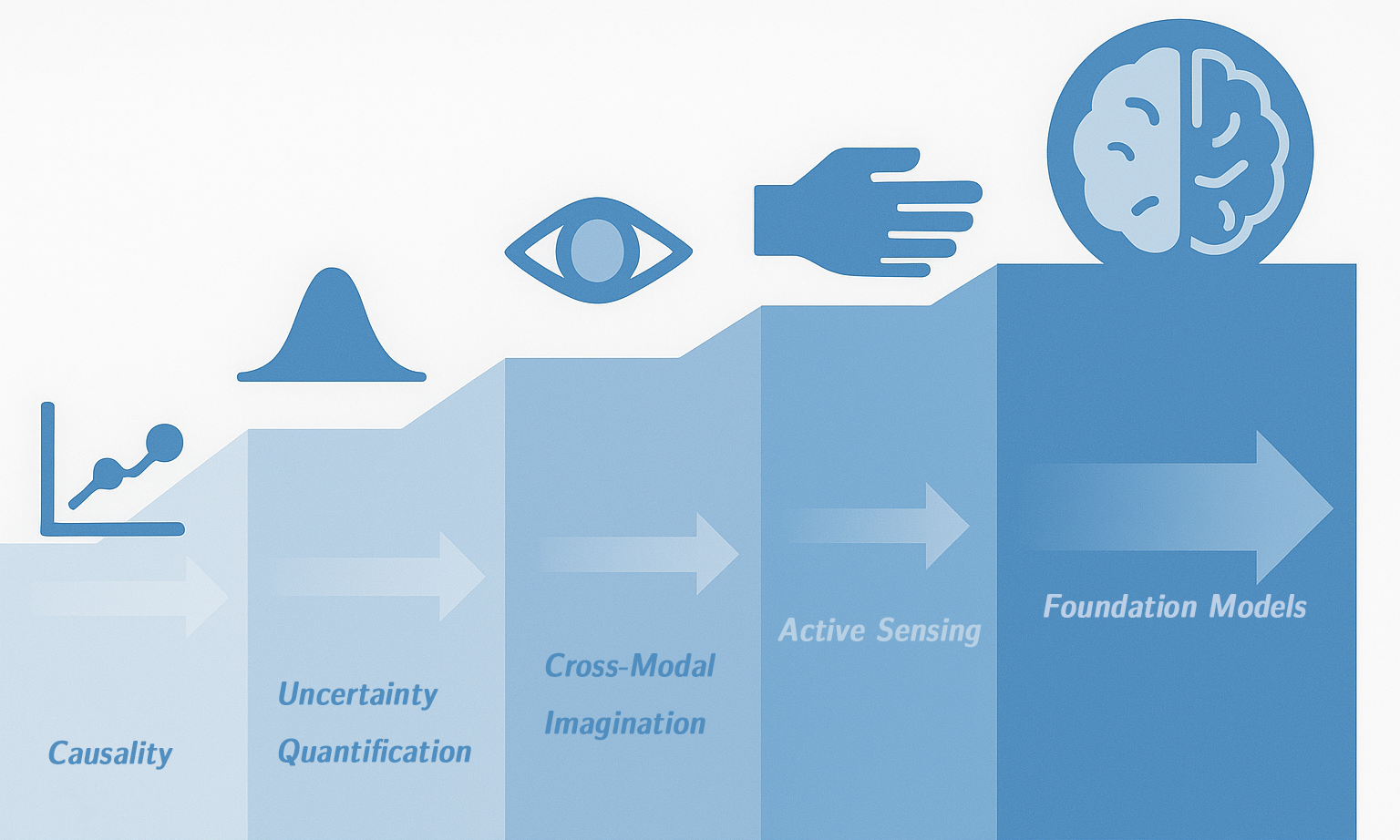}
    \caption{A proposed research roadmap for the next generation of acoustic world models. }
    \label{fig:roadmap}
\end{figure}

\textit{The roadmap illustrates a progression through five key research pathways: (1) Causality, moving from correlation to causal understanding; (2) Uncertainty Quantification, advancing from deterministic to probabilistic models; (3) Cross-Modal Imagination, achieving unified physical perception; (4) Active Sensing, shifting from passive listening to embodied interaction; and (5) Foundation Models, aiming for a general-purpose physical acoustic intelligence. The structure suggests how these pillars build upon each other towards more robust and generalizable AI.}

\section{Conclusions and Future Work}
This survey has presented a comprehensive, in-depth overview of the principles, methodological frameworks, and applications of world models built upon the rich foundation of acoustic physical information. Our central thesis posits a paradigm shift: moving beyond treating sound as a mere sensory channel for event detection, and instead embracing it as a direct, physical medium for understanding the world. We have established that unlike vision, which primarily captures surface-level phenomena susceptible to occlusion and illumination variance, acoustic signals are the radiated mechanical energy of physical events. Governed by the fundamental laws of elastodynamics and aeroacoustics, they carry intrinsic, latent "acoustic signatures" of material composition, internal geometry, contact dynamics, and fluid flow. This makes sound a powerful, penetrating modality for AI to develop a form of "physical intuition," enabling a deeper, more robust, and more causal grasp of its environment that is fundamentally complementary to other senses.

The pathway to unlocking this potential lies in the synergistic integration of three core methodological pillars. We demonstrated how PINNs provide the important "gray-box" framework, embedding physical laws as strong regularizers to solve ill-posed inverse problems and ensure physical plausibility, even with sparse data. Complementing this, generative forward models—from differentiable simulators to neural acoustic fields—act as the AI's "imagination," empowering it to predict future sensory inputs, synthesize physically-grounded audio for controllable worlds, and serve as powerful data factories for training. Finally, self-supervised and multimodal learning frameworks provide the scalability, learning from the natural co-occurrence of unlabeled sensory streams to build powerful, high-level representations that form the semantic bedrock for downstream tasks. It is the fusion of these approaches—the rigor of physics, the predictive power of generation, and the scalability of self-supervision—that forms the cohesive engine for acoustic intelligence.

The transformative impact of these acoustically-grounded world models is already materializing across a broad spectrum of high-stakes domains. In robotics, they promise to endow machines with haptic-like sensitivity, enabling delicate manipulation and robust navigation in visually-denied environments. For autonomous driving, they offer a vital, omnidirectional safety layer, detecting non-line-of-sight hazards and classifying road conditions with a fidelity that vision alone cannot guarantee. In healthcare and finance, these models are democratizing diagnostics by turning everyday devices into non-invasive health monitors and uncovering subtle, non-consensus alpha signals in vocal biomarkers that transcend traditional text analysis.

However, the path forward is paved with significant scientific, computational, and, most thoroughly, ethical challenges. As this survey has detailed, the power to listen is a dual-use capability. The pervasive nature of acoustic sensing demands a non-negotiable commitment to "privacy-by-design," robust on-device processing, and federated learning. Algorithmic bias in acoustic data can perpetuate and amplify societal inequities, mandating auditing and the development of fairness-aware algorithms. Moreover, in high-stakes decisions, the "black box" is unacceptable; the pursuit of explainable AI for acoustic models is not a mere academic exercise but a prerequisite for accountability and public trust.

Looking ahead, the research pathways outlined—towards causal inference, probabilistic uncertainty quantification, unified cross-modal imagination, active embodied sensing, and ultimately, a foundational model for physical acoustics—provide a clear and ambitious roadmap. This is not simply a technical agenda; it is a blueprint for evolving AI from a sophisticated pattern-matcher into a nascent scientific reasoner, capable of forming hypotheses, performing experiments, and understanding the world through its underlying physical laws. Continued, dedicated research and, above all, responsible development in this area hold the promise of building safer, more capable, and more humane artificial intelligence. By learning to listen to the physics of the world, AI can lay an important and enduring cognitive foundation for the future of human-machine symbiosis, where technology understands our world not just by its appearance, but by its very substance.

\par\vspace{1em}
In summary, the key contributions of this survey are:
\begin{itemize}
    \item It defines and frames the emerging field of world models grounded in acoustic physical information.
    \item It establishes the central role of acoustic signals in physical perception and causal reasoning, moving beyond surface-level correlations.
    \item It provides a structured review and comparative analysis of the core methodologies that fuse physical principles with advanced machine learning techniques.
    \item It details the significant real-world impact and future potential of acoustic world models across diverse, high-stakes application domains.
    \item It analyzes the primary technical and ethical challenges with deep case studies, and proposes a concrete, forward-looking research roadmap with specified technical hurdles and breakthroughs to guide future innovation.
\end{itemize}

\subsection*{Next Steps in Our Research}
Based on the framework and roadmap detailed in this survey, our own research group is actively pushing these frontiers. Our prior work has already laid a practical foundation for some of these concepts. For instance, in \cite{chen2025synergistic}, we introduced a novel framework that integrates nonlinear acoustic computing with reinforcement learning to enhance human-robot interaction under severe noise and reverberation. By embedding physically informed wave equations like Westervelt and KZK into a reinforcement learning loop, our system adaptively optimizes parameters such as beamforming and absorption, demonstrating superior noise suppression and accuracy over traditional methods.

Looking ahead, we plan to extend this line of inquiry in two primary directions. First, we will further develop the concept of the acoustic world model specifically for the challenging task of human vocal denoising in complex, dynamic scenes. Instead of treating this as a purely statistical signal separation problem, our goal is to build models that can physically reason about the environment—including room geometry, noise source characteristics, and multipath propagation—to "imagine" and reconstruct the clean vocal signal. Second, we will continue to deepen our exploration of these research ideas within the domain of robotics. The main goal is to move beyond passive sound perception and give robots a true physical understanding from sound, enabling more reliable navigation, safer human–robot teamwork, and finer object handling, thus advancing the human–machine partnership highlighted in this survey.

\subsection*{Limitations in Our Survey}
While we have strived for a comprehensive and structured overview, we acknowledge certain limitations inherent in this work. Firstly, our survey's primary focus is on the computational and machine learning frameworks for acoustic world models. Consequently, the equally important domains of advanced acoustic sensor design, microphone array signal processing, and the nuances of physical acoustics theory itself were discussed primarily in service of the AI models, rather than as standalone topics. A deeper exploration of the co-design of novel sensing hardware and learning algorithms represents a vital area for future reviews. Secondly, in balancing breadth with depth, our review of methodologies necessarily provides a high-level synthesis. A deep, mathematical dive into the intricacies and variants of every discussed algorithm (e.g., different PINN formulations or diffusion model architectures) was beyond our scope. Lastly, the field of AI is advancing at a breathtaking pace. This survey serves as a foundational snapshot at a specific point in time. We encourage readers to view it as a starting point and to actively follow the latest developments in top-tier conferences, as new state-of-the-art approaches are continually emerging. These identified gaps represent fertile ground for future research and more specialized reviews.

\section*{Acknowledgements}
The authors would like to express their sincere gratitude to the broader research community for their pioneering work, which formed the foundation of this survey. We are particularly indebted to the researchers and faculty at the Institute of Acoustics, Chinese Academy of Sciences, Peking University, and The University of Hong Kong for their foundational contributions to acoustics, signal processing, and artificial intelligence. Their work and the insightful discussions they have fostered within the academic community have been a profound source of inspiration. We also thank the anonymous reviewers for their constructive comments and suggestions, which have significantly improved the quality and clarity of this paper.
\appendix
\section{A List of Benchmarks, Datasets, and Simulators}

To facilitate further research and provide a practical starting point for newcomers to the field, this appendix lists key resources, including public datasets, simulation environments, and standardized benchmarks. These resources are foundational for developing, training, and evaluating acoustic world models.

\subsection{Public Datasets for Acoustic Perception and Modeling}
\begin{itemize}
    \item \textbf{AudioSet:} Developed by Google, this is one of the largest publicly available datasets for general sound event detection. It consists of over 2 million weakly-labeled 10-second video clips from YouTube, annotated with an ontology of 632 sound event classes. It is invaluable for pre-training large-scale audio representation models \cite{gemmeke2017audio}.
    \item \textbf{FSD50K:} An open dataset of human-labeled sound events containing over 51,000 audio clips drawn from the Freesound database. Its high-quality manual annotations and permissive license make it a go-to resource for supervised sound event detection and classification tasks \cite{fonseca2021fsd50k}.
    \item \textbf{WHAM!:} A dataset designed for speech separation tasks. It generates mixtures by combining speech from the WSJ0 corpus in simulated anechoic and reverberant rooms. Its provision of clean source signals, mixtures, and the corresponding room impulse responses (RIRs) makes it highly suitable for research on acoustic propagation and dereverberation \cite{wichern2019wham}.
    \item \textbf{TAU Acoustic Scenes Series:} A collection of datasets released as part of the DCASE (Detection and Classification of Acoustic Scenes and Events) challenge. These datasets are meticulously recorded in various real-world locations (e.g., streets, parks, public transport) using standardized equipment, making them a primary benchmark for acoustic scene classification \cite{mesaros2016tut}.
\end{itemize}

\subsection{Simulation Environments for Embodied and Physical AI}
\begin{itemize}
    \item \textbf{AI Habitat (Habitat-Sim):} A high-performance 3D simulator with an emphasis on photorealism and speed, primarily for training embodied AI agents (e.g., robots) \cite{savva2019habitat}. Its extension, Habitat-Sound, integrates physically-based sound rendering, enabling research on audio-visual navigation and acoustic mapping in complex indoor environments \cite{chen2020soundspaces}.
    \item \textbf{Isaac Gym / Isaac Sim:}  A suite of simulation tools from NVIDIA that leverage GPU acceleration for large-scale, physics-based reinforcement learning. Its strength lies in simulating rigid and soft-body dynamics, articulations, and contact mechanics, making it an ideal platform for generating physically accurate interaction sounds for robotic manipulation tasks \cite{makoviychuk2021isaac}.
    \item \textbf{ThreeDWorld (TDW):} A multimodal, interactive, and physically realistic simulation platform \cite{gan2020threedworld}. TDW features a sophisticated procedural audio engine (PyImpact) that can synthesize physically-based impact and contact sounds in real-time based on the objects' material properties, geometry, and collision dynamics. It is exceptionally well-suited for studying the causal link between physical interactions and their acoustic signatures.
    \item \textbf{SAPiEN:} A realistic and articulated object-centric simulator for robotics. It provides accurate physics simulation, which is important for tasks involving robotic manipulation and learning from interaction. It can serve as a powerful data generator for action-conditioned acoustic models \cite{xiang2020sapien}.
\end{itemize}

\subsection{Standardized Benchmarks and Challenges}
\begin{itemize}
    \item \textbf{DCASE Challenge:} The premier annual competition for the "Detection and Classification of Acoustic Scenes and Events". It provides well-defined tasks, public datasets, and a standardized evaluation framework, driving progress in a wide range of acoustic analysis problems \cite{mesaros2025decade}.
    \item \textbf{Audio-Visual Navigation Benchmark:} A common task within embodied AI challenges (e.g., at the Conference on Computer Vision and Pattern Recognition (CVPR)). The agent is placed in a 3D environment and must navigate to a sounding object using both visual and auditory inputs. It serves as a key benchmark for multimodal perception and action, heavily popularized by simulators like Habitat-Sound \cite{chen2020soundspaces}.
    \item \textbf{HEAR Challenge:} A benchmark for the "Holistic Evaluation of Audio Representations" designed to evaluate the generalization of learned audio representations. Models are pre-trained and then evaluated on a diverse suite of 19 downstream tasks, providing a comprehensive assessment of a representation's quality and robustness \cite{turian2022hear}.
\end{itemize}

\section{A Living Repository of Resources}
For a comprehensive and continuously updated list of resources, including direct links to datasets, simulators, seminal papers, and open-source codebases relevant to this survey, please visit our curated repository on GitHub:    \url{https://github.com/soundai2016/survey_acoustic_world_models}

\bibliography{references}

\end{document}